\documentclass[sigconf,anonymous=false,nonacm=true]{acmart}
\usepackage[utf8]{inputenc}
\usepackage{amsmath}

\setcopyright{rightsretained}

\renewcommand{\citeA}[1]{\citeauthor{#1}~\cite{#1}}

\begin{document}

\title[Tackling Racial Bias in Automated Online Hate Detection]{Tackling Racial Bias in Automated Online Hate Detection: Towards Fair and Accurate Classification of Hateful Online Users Using Geometric Deep Learning}

\author{Zo Ahmed}
\affiliation{%
  \institution{Oxford Internet Institute,\\University of Oxford}
  \city{Oxford}
  \country{United Kingdom}
}

\author{Bertie Vidgen}
\affiliation{%
  \institution{The Alan Turing Institute}
  \city{London}
  \country{United Kingdom}}
\email{bvidgen@turing.ac.uk}

\author{Scott A.\ Hale}
\affiliation{%
  \institution{Oxford Internet Institute,\\University of Oxford\\Meedan}
  \city{Oxford}
  \country{United Kingdom}
\email{scott.hale@oii.ox.ac.uk}
%\orcid{1234-5678-9012}
}

%%
%% By default, the full list of authors will be used in the page
%% headers. Often, this list is too long, and will overlap
%% other information printed in the page headers. This command allows
%% the author to define a more concise list
%% of authors' names for this purpose.
\renewcommand{\shortauthors}{Ahmed, Vidgen, \& Hale}

%%
%% The abstract is a short summary of the work to be presented in the
%% article.
\begin{abstract}
Online hate is a growing concern on many social media platforms and other sites. To combat it, technology companies are increasingly identifying and sanctioning `hateful users' rather than simply moderating hateful content. 
Yet, most research in online hate detection to date has focused on hateful \emph{content}. This paper examines how fairer and more accurate hateful user detection systems can be developed by incorporating social network information through geometric deep learning. Geometric deep learning dynamically learns information-rich network representations and can generalise to unseen nodes. This is essential for moving beyond manually engineered network features, which lack scalability and produce information-sparse network representations.
This paper compares the accuracy of geometric deep learning with other techniques which either exclude network information or incorporate it through manual feature engineering (e.g., node2vec). It also evaluates the fairness of these techniques using the `predictive equality' criteria, comparing the false positive rates on a subset of 136 African-American users with 4836 other users.
Geometric deep learning produces the most accurate and fairest classifier, with an AUC score of 90.8\% on the entire dataset and a false positive rate of zero among the African-American subset for the best performing model. This highlights the benefits of more effectively incorporating social network features in automated hateful user detection. Such an approach is also easily operationalized for real-world content moderation as it has an efficient and scalable design.

\end{abstract}

\maketitle

\section{Introduction}

% online hate is a thing and inflicts real harm on people
The massive expansion in online social media over the last two decades has brought unprecedented connectivity and communication to people. Unfortunately, such communication is sometimes characterised by harm and abuse, such as hate directed against groups. This hate not only negatively impacts users of online platforms and their communities \cite{Sabatini2017}, it can also stir up social tensions and affect the reputation of the platforms who host them \cite{Statt2017}. Online hate can also have implications in the offline world: it has been linked to youth suicides, mass shootings, violent hate crimes, and extremist recruitment \cite{Johnson2019}. 

% detecting hate is hard -- focusing on users is important
The vast complexity and scale of online ecosystems means that hate can only be scalably, efficiently, and accurately detected through the use of computational and statistical techniques, such as machine learning (ML). Hateful \emph{content} detection has received substantial academic attention but less attention has been paid to hateful \emph{user} detection. This is a notable omission given a growing body of research which shows that much online hate is sent by `hateful users’, who share a distinct set of linguistic features, personality and activity traits, and network characteristics \cite{RibeiroA, qian-etal-2018-leveraging}. For instance, a 2017 study by Demos found that of a corpus of Islamophobic tweets they identified, 50\% were sent by just 6\% of users \cite{Islamophobia2019}; targeting these users is a promising way of maximising resources to tackle online hate. Social media platforms are increasingly aware of this and have adapted their moderation efforts accordingly. For example, since 2017, Twitter has regularly changed its rules to more stringently suppress hate, suspending entire accounts which may be hateful instead of simply taking down violating content \cite{Twitter2019}. Several prominent purveyors of hate, such as anti-feminist writer Robert Stacy McCain, alt-right racist Milo Yiannopoulos, and British ultra-nationalist Tommy Robinson, have since been banned.

% what we are doing
This paper examines how different approaches to incorporating network information affect both the accuracy and fairness of machine learning models to detect hateful users. We use a partially annotated retweet network dataset containing up to 200 tweets per user for 100,000 unique users \cite{RibeiroA} and find that geometric deep learning results in both fairer and more accurate hateful user detection compared to linear regression, support vector machines, and artificial neural networks with user and manually-engineered network features (e.g., network centrality measures). This approach also has the benefit that it is far more scalable and efficient for use in real-world settings.

% Structure
In Section \ref{sec:background}, we explain the benefits of incorporating social network features by reviewing research on the social dynamics of hateful behaviour online. We also appraise existing automated approaches for hate detection, highlighting their context-blindness and the challenges they face in producing fair and accurate outcomes. We then review the concept of fairness in machine learning, and the particular issues of bias for automated hate detection.
In Section \ref{sec:data} and Section \ref{sec:methods}, we outline the data and methods, describing the dataset we use and methods for detecting hateful users and identifying African-Americans.
In Section \ref{sec:results}, we present and compare the results of various classification models in terms of their overall performance as well as their performance on the subset of African-American users. We also conduct an error analysis to better diagnose the best-performing model's performance, and identify limitations.
In Section \ref{sec:discussion}, we discuss the results and their implications for the accuracy and fairness of hateful user detection, as well as limitations of the research.
% Lastly, in Section \ref{sec:conclusion} we suggest fruitful areas for future research.

\section{Related work}\label{sec:background}

\subsection{The social dynamics of online hate}

Online hate is a contested and context-dependent concept, and despite growing concerns about its harmful effects, platforms, governments and researchers have been unable to reach a common definition \cite{vidgen2019challenges}.
Academic work displays a fragmented understanding of online hate \cite{MacAvaney}, although most definitions share three common elements: content, intent, and harm \cite{Marwick2014}. Content relates to the textual, visual or other modes used to proliferate hate. Intent and harm are more difficult to discern, and relate to the thoughts and experiences of the perpetrator and victim \cite{Pitsilis2018,Olteanu2017}.

Social psychological work on prejudice and abuse proposes that the expression of hate is strongly influenced by intergroup dynamics, such as contact and conflict between competing groups \cite{Christ2014}. Following Crandall and Eshelman's Justification--Suppression model of prejudice, \citeA{Day2018} finds racism on Twitter flourishes when the offender is socially motivated by a group of like-minded racists, when prejudice against targets is easy to socially justify, and when capable guardians are not present to suppress their activities. It has been suggested that similar dynamics operate amongst hateful far-right online communities \cite{Eckstrand2018}.

Hateful users are often homophilous and organise themselves into densely connected and clustered networks where hateful ideas and content proliferate quickly \cite{RibeiroA, Johnson2019}. \citeA{Johnson2019} argue that content removal proves largely ineffective in such settings: the sources of the content---hate-spewing, highly resilient clusters of users who are committed to their goals---live on, growing larger with time due to consolidation of smaller groups or attraction of new recruits. In line with these findings, platforms are increasingly seeking to identify and remove hateful users rather than content. Network features, which have been effective for predicting other attributes \cite{Mislove2010}, have also been shown to assist with detecting hateful users \cite{Mishra2018, RibeiroA}

\subsection{Automated methods of online hate detection}
Barring a handful of papers, most academic research on detecting online hate has focused on content, rather than the users behind it. %This is despite the fact that the two tasks are highly related: incorporation of user-level features, though rare, has been successfully used to boost the performance of online hate detection systems \cite{unsvag2018}, while detection of hateful users usually leverages the information in their content (indeed, this is typically how they are identified as 'hateful').
Incorporating context has been shown to improve hateful \emph{content} detection, with several papers deploying transformer-based models such as BERT to better distinguish between hateful and non-hateful content, even when they have lexical similarities \cite{MacAvaney}. However, despite improving performance overall, such models are often still unable to distinguish the context of polysemous words related to minority groups \cite{Mozafari2019}. Part of the challenge is that the meaning of such terms depends on the identity of the speaker and how they are used (e.g., a black person using the term ``N*gga'' is fundamentally different to a white person). With short text statements (e.g., tweets) it can be difficult to capture and represent these different uses (particularly if the starting dataset is small), even with a transformer-based model.

Previous research shows that user-level data can be leveraged to improve classification accuracy, especially where text alone does not offer adequate signals. This includes both manually engineered network features such as in- and out-degree as well as shallow-embedding methods such as node2vec \cite{Chatzakou2017, Mishra2018}. 
\citeA{unsvag2018} create a classifier in which textual features are augmented with user-specific ones, such as gender, activity and network (number of followers and friends). They use Logistic Regression (LR), reporting a 3 point increase in the F1 score with the addition of user-level features.
\citeA{Mishra2018} use a variant of the word2vec skipgram model by \citeA{Mikolov2013} to create network (node2vec) embeddings of users based on their network positions and those of their neighbours. They concatenate the generated user network vectors with character n-gram count vectors, and use a LR classifier, achieving a 4 point increase in the F1 score compared to the same model without the network embeddings.

One drawback of node2vec embeddings, however, is that they can only incorporate the structural and community information of a user’s network; the textual content of tweets composed by other users in this network is not incorporated. This may, as \citeA{Mishra2019-gcn} demonstrate, lead to misclassification of a normal user’s content because of their hateful network neighbourhood through `guilt-by-association'.
Most importantly, from a practical perspective, they may be unfeasible as the embeddings need to be re-generated for the entire graph every time a new user enters the network \cite{Hamilton2017}. The large user base of social media platforms renders such methods computationally impractical.

Moving from detecting hateful content to detecting hateful users is not straightforward due to challenges in sampling \cite{Davidson2017, Vidgen2018, RibeiroA} and in accessing network and other user attributes \cite{Jia2017}.
Geometric deep learning offers a very promising avenue for hateful user detection.
\citeA{RibeiroA} were the first to use geometric deep learning for this task, using GraphSAGE, a semi-supervised learning method for graph structures such as networks. It incorporates a user’s neighbourhood network structure as well as the features of her neighbours to learn node embeddings for unseen data \cite{Hamilton2017}.
%Geometric methods leverage the whole spectrum of textual, user-specific and network-level features in a learnable fashion for prediction.

\citeA{RibeiroA} achieve an overall unbalanced accuracy of 84.77\% and F1-score of 54\%  but they leave some unanswered questions that this study fills.
While they find that including learned network embeddings improves accuracy substantially, their baseline comparison model is a decision tree which lacks any network information. With one model having access to more features, such comparisons are problematic as they are not truly like-for-like. Moreover, they fail to compare their model to different types of geometric deep learning or against deep learning methods. Lastly, they fail to consider the implications of such methods beyond merely the accuracy of their outcomes: network-level features may, theoretically at least, be less biased towards the `guilt-by-association' problem because of how they exploit homophily in social networks. This is a substantial limitation which needs to be addressed if social network information is to be fairly incorporated into hateful user detection models.

\subsection{Fairness in automated hate detection}
Attention in online hate detection research has increasingly shifted from solely considering \textit{performance} to also evaluating models based on their \textit{fairness}.
Numerous groups can be victims of unfair and biased treatment in online settings, including women, LGBTQ people, and racial minorities. We focus on the experience of African-Americans as this is particularly relevant for hateful user/content detection.

\citeA{Chung2019} show that Google’s Perspective toxicity classifier disparately impacts African-American users by disproportionately misclassifying texts in African-American Vernacular English (AAVE) as toxic. Exclusively learning from textual elements, the model identifies relationships between words such as `black’, `dope’, or `ass’ and toxicity, as such words appear frequently in hateful posts. Given the frequent use of such words in AAVE, African-American users bear the negative implications of such erroneous learning \cite{Chung2019}. 

\citeA{Davidson2019} use machine learning to identify AAVE tweets %\cite{Blodgett2016} 
and find substantial and statistically significant racial disparity in the five datasets they investigate. AAVE tweets are, for some datasets, more than twice as likely to be associated with hate than Standard American English (SAE) tweets \cite{Davidson2019}. They investigate how tweets containing `nigga’, a term re-appropriated by the African-American community but still frequently used to perpetrate hate, are classified. They find that a tweet’s dialect, independent of whether it uses `nigga', strongly influences how it’s classified: AAVE tweets are more likely to be classed as hate-related than SAE tweets, even where both contain the term `nigga' \cite{Davidson2019}.

\citeA{Davidson2019} argue that annotation errors in the training data, where black-aligned language may be more likely to be labelled as hateful, may offer an additional explanation. \citeA{Sap2019} have confirmed this possibility, exploring two key hate speech datasets and finding that AAVE tweets are up to two times more likely to be labelled as offensive by human annotators compared to non-AAVE tweets. Models trained on such datasets learn and propagate these annotation biases, amplifying disparate impact. Even BERT-based classifiers can confuse the meaning of words such as `women', `nigga' or `queer', falsely flagging non-hateful content that uses such words innocuously or self-referentially \cite{Mozafari2019}. 

Thus far, these issues of fairness in hate detection largely been explored in relation to \textit{content} but are likely to also affect hateful \emph{user} detection algorithms. Yet, to our knowledge, they have not been explored to date. It is crucial that the gap between current academic focus and the wider need for fair and accurate hateful user classification is filled, especially given the increased use of such systems by social media platforms. 

Fairness and bias are contested concepts in the wider ML literature.
\citeA{Mehrabi2019} provide a comprehensive review of work on `Fair ML', and define fairness as ``the absence of any prejudice or favouritism towards an individual or a group based on their intrinsic or acquired traits in the context of decision-making". However, they acknowledge that fairness can have varying notions both within and across disciplines, depending on how it is operationalized.
\citeA{Barocas2017} have categorised notions of fairness into two groups: those centred around `disparate treatment’ and those around `disparate impact’. Disparate treatment characterises decision-making if a subject’s sensitive attributes, such as race, partly or fully influence the outcome. In contrast, disparate impact occurs where certain outcomes are disproportionately skewed towards certain sensitive groups \cite{Barocas2017}.
Elsewhere, \citeA{Kleinberg2016} show the constraints imposed by different definitions of fairness on ML systems are incompatible: in most practical applications ML models cannot optimise for a particular definition of fairness without also causing unfair outcomes when considered from a different definition. Thus, it is important to establish the notion of fairness that is being prioritized before models are optimized.

At the same time, fairness constraints may lead to a `fairness-accuracy trade-off', where ``satisfying the supplied fairness constraints is achieved only at the expense of accuracy" \cite{NIPS2019_9082}.
However, \citeA{NIPS2019_9082} have also shown that theoretically and practically, such a trade-off may be overcome in cases where systemic differences in the outcome variable (e.g., online hate) are not caused by the sensitive attribute (i.e., race). This is a plausible assumption in the current setting: nothing suggests that people of a particular race are intrinsically more hateful than those of another, especially after accounting for other factors. As such, it could be possible for both fairness and accuracy to increase, given an effective model.

Fairness should be defined based on the context and nature of the task, and bearing in mind the trade-offs involved \cite{Mehrabi2019}. One notion of fairness is demographic parity, which requires that a model's predictions are the same across different sensitive groups \cite{Hardt2016}. While applicable in certain contexts, this may be suboptimal for hate detection tasks. The social reality of online hate means that certain groups are more likely to perpetrate it, such as white supremacists compared with other users (of any race) \cite{Davidson2019}. However, because demographic parity does not incorporate accuracy of classifications (i.e., whether predicted outcomes match actual labels), it could be `achieved' even when such users are misclassified as normal. Secondly, the constraint of statistical parity across groups may actually add to the unjust penalisation of minority groups, such as African-American users: they are usually victims of hate, but to equalise outcomes across groups, a particular ratio of such users will have to be classified as hateful. As such, demographic parity is not used as an evaluation criterion in this paper. 

Instead, we use the notion of `predictive equality’ (a.k.a false positive error rate balance) to evaluate fairness \cite{Verma2018}. Predictive equality is very appropriate to studying online hate, which has a history of disproportionate false positives among minority groups. There is also a magnified harm of such errors compared to false negatives, given that it penalises the very groups that hate classifiers are meant to protect \cite{Dixon2018}. Bias in hate detection has been shown to occur mostly at the group level \cite{Sap2019}.
In practice, this means that the false-error rate for African-Americans should be no higher than that for all other users.

% As defined by \citeA{Verma2018}, a binary predictor's prediction $d$ satisfies predictive equality with respect to $G$ (the protected attribute) and $Y$ (the outcome), if  
% \begin{equation}
% P\left(d=1 \middle| Y=0, G=m\right)=P\left(d=1 \middle| Y=0, G=f\right)
% \end{equation}
% . Here f and m may be two different sensitive group attributes, for example white and African-American. Thus, the probability of a person in the negative class (here, the normal category) being falsely classified as hateful should be the same across groups \cite{Verma2018}. 

Even with an appropriate fairness definition, combating bias is a complex and multifaceted task, where methodological innovation is required. 
Some existing strategies, such as race priming during annotation \cite{Sap2019, Moskowitz2011} are limited in effectiveness, often suffering from the need for annotators to guess a user's race or other demographics \cite{Lahoti2020}. Moreover, even where such data is available, \citeA{Moskowitz2011} find that race priming may only have its desired effect of stereotype suppression if a person already has egalitarian goals in the first instance. Thus, the impact on annotations of race priming depends primarily on the character of the annotators, rather than the information they receive.

Strategies to mitigate bias through more context-aware models have also been proposed (e.g., using transformer-based models) but they have mostly been geared exclusively towards \textit{content} classification, sidelining the issue of hateful user detection. Despite using contextual word embeddings, for example, \citeA{Mozafari2019} find that their classifier still fails to capture the meaning of identity-related words such as `woman' or `queer', classifying tweets using them as hateful.

% \citeA{Vaidya2019} propose interventions to the model architecture itself: seeking to reduce false positives among minority classes, they develop a multi-task learning framework which simultaneously classifies toxicity and identifies information from text, using an attention layer to help the model effectively capture textual meaning while incorporating the contextual information provided by user identity.
% They use data from Kaggle’s Unintended Bias in Toxicity Classification Challenge, which includes 1.8 million comments along with information on their toxicity and the possible identity of their author. Compared to Google’s own Perspective algorithm, such a model predicts 10\% less toxicity for non-toxic comments that mention usually hateful words self-referentially, for example ``I am a gay man”. This advance is noteworthy, but is still characterised by a key defect: the dataset used has identity features which are guessed by annotators from the text. 
% This may be highly prone to error, given that societal stereotypes on minority-group dialects often result in erroneous labelling \cite{Davidson2017}. 

An effective and relatively low-cost way of incorporating user-specific information to improve fairness may involve leveraging users' social networks. Given homophily in such networks, network-incorporating ML models can learn user characteristics without these being manually specified. Indeed, new approaches to `Fairness without Demographics' in ML have sought to exploit correlations between observed features and unobserved sensitive group attributes to boost fairness \cite{Lahoti2020}. Homophily in race may be exploited in a similar vein for more context-aware and fairer automated hate detection. As \citeA{Wimmer2010} find, such homophily is prevalent in social media: African-American individuals, for example, are more likely to associate themselves with other African-Americans on social media platforms, creating a ``high degree of racial homogeneity" in social networks \cite{Wimmer2010}. When these networks are effectively incorporated into ML models, this homogeneity can be exploited to infer contextual information around an individual's identity, without directly extracting and feeding such features into the model. This may not only lead to more scalable models for online hate detection, but also fairer and more accurate ones. The text of a user can be contextualised against their identity, helping resolve ambiguities stemming from the analysis of text alone, such as with polysemous words like `nigga'.
%Moreover, such research should also focus on automated hateful \emph{user} detection, whose fairness is yet to be systematically evaluated.

\section{Data}\label{sec:data}

The data used in this paper was prepared by \citeA{RibeiroA} and obtained from Kaggle. It consists of a Twitter retweet network, with a total of 100,386 users (nodes), and 2,286,592 edges (including self-loops) between them. We removed all self-loops from the dataset.

The network is a directed graph $G = (V, E)$. Each node $u \in V$ represents a user, and a directed edge between two users $(u_i, u_j) \in E$  denotes that $u_i$ has retweeted $u_j$. As such, the edge directions can serve as a proxy for the direction of influence flow in the graph \cite{RibeiroA}. At the first step, \citeA{RibeiroA} use a Directed Unbiased Random Walk (DURW) algorithm to construct a sampled graph of all Twitter users.  DURW avoids oversampling high-degree nodes, producing a more representative sample \cite{Ribeiro2012}. The starting node is set to a Twitter account geo-located in San Jose, California.

\citeA{RibeiroA} selected 4,972 users for manual annotation. The relative sparsity of hateful online content has often led researchers to follow a lexicon-based approach to bias sampling towards collecting more hateful content \cite[e.g.,][]{Davidson2017, Burnap2015}. However, such approaches can bias sampling towards specific users who heavily use words in the specified lexicon; indeed the entire dataset of 1,972 racist tweets collected by \citeA{Waseem2016} came from only 9 users. To  mitigate this,  \citeA{RibeiroA} used a graph diffusion process based on a lexicon of words mainly used to express hate, taken from the Hatebase.org and ADL databases. The graph diffusion process assigns probabilities to each user based on their tweets and those of their nearest neighbour. It ensures a more balanced sample between hateful and innocuously used words in the lexicon as well as inclusion of users whose neighbours are hateful making it more likely that coded or non-standard hate is also included. % \cite{Magu2017}.

\citeA{RibeiroA} use crowdsourcing to manually annotate 4,972 of the sampled users as hateful or normal based on their sampled tweets and in reference to Twitter’s guidelines on hateful conduct. %which stated: ``You may not promote violence against or directly attack or threaten other people on the basis of race, ethnicity, national origin, caste, sexual orientation, gender, gender identity, religious affiliation, age, disability, or serious disease. We also do not allow accounts whose primary purpose is inciting harm towards others on the basis of these categories." \cite{Twitter2019}.
For each user, 3 independent annotators were given; if there was any disagreement then up to 5 annotators were assigned \cite{RibeiroA}. 544 hateful (10.9\%) and 4,438 normal (89.1\%) users were identified. 

For each user, \citeA{RibeiroA} extracted textual, user-level and network-level features. The text of users' tweets are represented as 300-dimensional GloVe vector embeddings, which were provided with the dataset. Other textual features were also derived, including sentiment and subjectivity scores. User-level features relate to their Twitter activity and profile, and include the number of followers, followees, statuses, and favourites normalised by the time since account creation. Manually-engineered network-level features include the in-degree, out-degree,  eigenvector, and betweenness centrality of each user, and aggregated user-level and textual attributes of their immediate 1-hop neighbourhoods. \citeA{RibeiroA} find large and statistically significant differences existed between normal and hateful users on such measures, with hateful users being identified as more active, central, and with newer accounts, likely because of previous account suspensions.
In addition, we retrieved the original tweets for a partial set of the users, which were provided to us by the original authors. We used these to examine classification errors and for demography detection but not for the modelling as they were not available for all users.

\section{Methods}\label{sec:methods}
The research methods used in this paper can be divided into two main categories: those for inferring users' demographics, and machine learning techniques for the automatic detection of hateful users. %Evaluation of the latter is conducted with respect to each research question, and involves a comparative investigation into various performance measures across techniques (for accuracy) and across demographic groups (for fairness). A detailed error analysis is also performed in relation to both questions.

\subsection{Detecting African-American users using a mixed-membership demographic language model}
Given the size of social media datasets, large-scale statistical tools have been developed to automatically infer demographic attributes, which can be used to diagnose and mitigate potential bias in the models deployed on such data \cite{Wang2019}. A subset of such research is devoted to inferring the race or ethnicity of social media users based on their content, profiles, or both \cite[e.g.,][]{Blodgett2016, Preot2018}. For this paper, a distantly supervised probabilistic topic model developed by \citeA{Blodgett2016} is used. It models dialect variation across races in the US \cite{Green2002}, using Latent Dirichlet allocation (LDA) to classify users into the four main races of the US Census: non-Hispanic white, non-Hispanic black, Hispanic, and Asian.
%Post-hoc analysis shows that the model accurately places high probabilities for the African-American category by effectively recognising the phonological and syntactical particularities of African-American English \cite{Blodgett2016}.
Using this model, \citeA{Blodgett2016} report accuracy of 97\% (AUC) for detecting users' race in a dataset of 26,009 users. It is appropriate for this research as the model was trained, at least partly, on US Census data and should work most effectively for US-based populations. Given that the random-walk induced strategy for creating the retweet network used here had as its starting seed node a user in San Jose, California, our sample of users is likely to mostly be from the USA \cite{hale2014global}. 

Following a similar approach to \citeA{Preot2018}, we applied the pre-trained topic model of \cite{Blodgett2016} to the message of each user, averaging the probabilities to determine a score for each user. If the average probability for the African-American category was above 0.8 we labelled them as African-American. At this first step, 168 users were identified with this approach; further manual verification was conducted at a second step, after which 32 users were identified as false positives, leaving 136 African-American users in the dataset. Of these, 8 are also annotated as hateful. 

\subsection{Classifying hateful users using machine learning}
In total, we train 7 models, 4 of which incorporate network information in different ways. 5-fold cross-validation is used to test the predictive power and fairness of the models, with a 80/20 train-test split. All models are coded in Python, using the TensorFlow2 library and the code is publicly available.
%\footnote{\emph{Removed for review}}

\subsubsection{Baseline classifiers}
Four baseline classifiers were trained, against which the performance of graph neural networks (GNNs) are compared on both accuracy and fairness: logistic regression, Support Vector Machine (SVM) with a radial basis function kernel, and two fully connected Artificial Neural Networks (ANNs). One of the ANNs is trained on textual, user-level, and manually engineered network-level features described above; all other baseline classifiers are trained only on textual and user-level features. %The ANN, even when trained on just textual and user-level data, performs best among the baseline models. As such, network-level features are added to its input data to compare the impact of such features with richer network-level features which GNNs dynamically learn.

\subsubsection{Geometric deep learning and GraphSAGE}
Analysis of properties of edges, nodes and topologies of such networks has been subject to various statistical and social methodologies \cite[see][]{Kolaczyk2009}. However, predicting such properties using machine learning had been restricted to using shallow representations or encodings of such properties, such as through node2vec \cite{Grover2016} or TransE \cite{Bordes2013}. Such shallow encoders, however,  cannot generalise to unseen nodes. The entire graph must be first embedded before prediction can be performed \cite{Leskovec2019}. This makes them impractical for use on most online social networks, where new users are added to the network in a continual stream. Moreover, they incorporate only the graph's topology, ignoring attributes of the nodes themselves, such as a user's text or activity patterns in the current case \cite{Leskovec2019}. 

Geometric deep learning, through the use of GNNs, overcomes such difficulties, resulting in more information-rich and predictive node embeddings fine-tuned to a given task \cite{Wu2020}. Spatial Graph Convolutional Neural Networks (GCNNs), a variant of which is used here, create node embeddings by defining graph convolutions upon a node's spatial relations: each node's representations are obtained by convolving its embedding with those of its neighbours, leading to information propagation across a graph's edges while capturing its topological structure simultaneously \cite{Wu2020}. A spatial approach was chosen, instead of the also popular spectral approach, due to their better efficiency, more effective generalisability to new graphs, and capacity to incorporate directed edges \cite{Wu2020}. The latter two are especially relevant for constantly evolving and directed social networks such as Twitter.

We use GraphSAGE, an inductive spatial variant of GCNNs proposed by \citeA{Hamilton2017}. As opposed to learning distinct embeddings for each node, in what would be a `transductive' approach, GraphSAGE trains ``a set of aggregator functions that learn to aggregate feature information from a node's local neighbourhood", with each such function aggregating information from a different number of hops away from the target node in the network \cite{Hamilton2017}. The learnt parameters of these functions, shared across nodes in a form analogous to vanilla convolutional neural networks, are then used to generate embeddings for previously unseen nodes. Such an `inductive' or end-to-end approach learns both the topological structure of node neighbourhoods as well as the optimal aggregation of node features within them \cite{Hamilton2017}. 

GraphSAGE can incorporate different aggregation functions, and we use three GraphSAGE models with different aggregation functions: mean aggregation, which takes a weighted average of the embeddings of a node's neighbours; max pooling, which tranforms these embeddings through a neural network and takes the maximum activation; and self-attention, which uses a single attention layer to compute attention coefficients for a node's neighbours, better incorporating their differing importance for the target node's representation.

Other techniques such as vanilla Graph Convolutional Networks (GCN) or Graph Attention Networks, which uses multi-headed attention at each layer, are not used because of their high computational intensity: both have a complexity of $O(m)$, where $m$ represents the number of edges in the graph \cite{Wu2020}. GraphSAGE, meanwhile, has a time complexity dependent on the number of samples and layers; these can be tuned to reduce computational intensity, though some predictive power may be traded-off as a result \cite{Wu2020}. Despite being experimented on an NVIDIA Tesla M60 GPU with 2,048 parallel processing cores, the other techniques were too time-consuming on the network of +2M edges, and were subsequently not employed.

\section{Results}\label{sec:results}

\begin{table*}[t]
\centering
%\captionsetup{justification=centering}
\label{tbl:accuracy}
%\begin{tabular}{p{3.8cm}p{1.5cm}p{1.5cm}p{1.5cm}p{1.5cm}p{1.5cm}} 
\begin{tabular}{lrrrrr} 
\toprule
Model                & Accuracy        & Precision (\%)  & Recall (\%)     & F1-Score (\%)   & AUC (\%)        \\
% \multicolumn{1}{c}{Model}  & Accuracy (\%)~ & Precision (\%) & Recall (\%)   & F1-Score (\%) & AUC (\%)       \\ 
\midrule
$LR$                     & 83.1           & 37.6           & 82.1          & 51.6          & 89.9           \\
$SVM$                      & 81.6           & 34.8           & 77.6          & 48.0          & 87.6           \\
$ANN_{text+user}$          & 84.1           & 40.6           & 75.7          & 51.7          & 88.2           \\
$ANN_{text+user+network}$  & 87.2          & 45.3           & 68.8          & 54.2          & 87.2           \\
$GraphSAGE_{maxpool}$      & 84.2           & 40.3           & 80.9          & ~53.4         & 90.2           \\
$GraphSAGE_{attention}$    & 84.8           & 40.6           & \textbf{82.3} & 54.3          & 90.8           \\
$GraphSAGE_{meanagg}$      & \textbf{87.4}  & \textbf{46.1}  & 76.8          & \textbf{57.5} & \textbf{90.8}  \\
\bottomrule
\end{tabular}
\caption{Accuracy Metrics (Maximum columnar values in bold)}
\label{table:results-accuracy}
\end{table*}

\begin{table*}
\centering
%\captionsetup{justification=centering}
%\begin{tabular}{p{3.8cm}p{1.5cm}p{1.5cm}p{1.5cm}p{1.5cm}p{1.5cm}}
\begin{tabular}{lrrrrr} 
\toprule
Model                & Accuracy (AA users) (\%) & FPs (AA users) & FPs (non-AA users) & FP rate  (AA users)  & FP rate (non-AA users)      \\
\midrule
$ANN_{text+user}$ & 80.9  & 26  & 660 & 20.3\%  & 14.8\%  \\
$ANN_{text+user+network}$  & 86.0 & 13  & \textbf{465} & 10.2\% & \textbf{10.5\% }  \\
$GraphSAGE_{maxpool}$      & 91.7  & 5 & 680 & 3.9\% & 15.7\% \\
$GraphSAGE_{attention}$    & 93.9 & 1 & 662  & 0.7\%  & 15.3\%  \\
$GraphSAGE_{meanagg}$      & \textbf{94.2} & \textbf{0} & 492 & \textbf{0.0\%} & 11.1\% \\
\bottomrule
\end{tabular}
\caption{Fairness Evaluation Metrics \\ (African-American denoted as AA, False positive denoted as FP)}
\label{tbl:fpr}
\end{table*}

\subsection{Classification Accuracy and Fairness}

Table \ref{table:results-accuracy} shows the performance of each model. In terms of binary accuracy as well as F1-Score, models which incorporate network-level features perform better than those which do not, while neural-network based approaches outperform more traditional machine learning approaches. Geometric deep learning performs best across all metrics. The GraphSAGE model (using mean aggregation) has the best F1-Score (57.5) and AUC (90.8). 

Our results show the power of Geometric deep learning for achieving higher performance in hateful user detection.
Firstly, the logistic regression model, the simplest baseline used, still performs fairly well, even surpassing the ANNs on certain metrics.
Secondly, when incorporating network information into the ANN, its performance declines considerably on some metrics such as Recall (-6.9) and AUC (-1). However, these declines do not occur for the GraphSAGE models, which also incorporates network-related information. 
Lastly, all models have relatively high recall but low precision, indicating that they can detect hateful users well but often mistake normal users for hateful ones. The fact that AUC is still high, despite this, indicates that this disparity may not be entirely related to trade-offs made by the models between true positives and false positives. Instead, the size and unbalanced nature of the dataset is likely to have played a role (despite our use of class weights). Indeed, as \citeA{Juba2019} report through an investigation of 4 imbalanced datasets, ``in the presence of severe class imbalance, it is not possible to reliably achieve high precision and recall unless one possesses a large amount of training data". Method such as class weighting, over-sampling, or under-sampling are minimally effective and face an ``inherent statistical barrier" \cite{Juba2019}.

%\subsection{Fairness Evaluation}

Table \ref{tbl:fpr} shows the accuracy of each model on the subset of 136 African-American users, and compares the false positive rates (FPRs) for African-American and non-African-American users.
The results show that, firstly, geometric deep learning outperforms the neural network classifiers in producing the lowest FPR among African-American users, with the mean aggregation based GraphSAGE model bringing this rate to zero.
Secondly, the neural network without network information (ANN\textsubscript{text+user}) is biased against African-American users, producing a FPR that is almost 6 percentage points higher among these users compared to the non-African-American group. Incorporation of network information in the ANN helps lower and roughly equalise these rates. When such network information is dynamically learnt, through the GraphSAGE models, the FPR for African-American users drops even further, becoming much lower than that of the non-African-American group.

\subsection{Error Analysis}
\begin{figure}
\centering
%\captionsetup{justification=centering}
\includegraphics[width=\columnwidth]{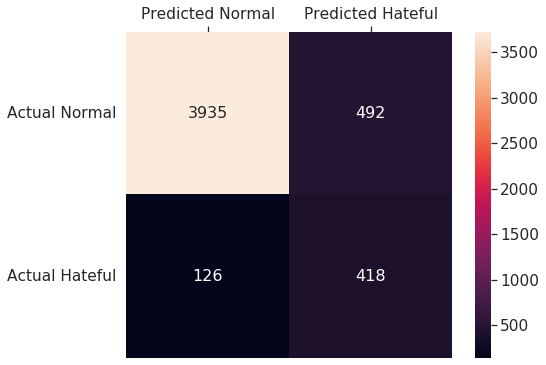}
\caption{Confusion matrix of predictions from best performing model: GraphSAGE with mean aggregation}
\label{fig:confusion-matrix}
\end{figure}

Figure \ref{fig:confusion-matrix} shows the classification confusion matrix for the best performing model, GraphSAGE with mean aggregation. As shown by the performance metrics reported above, the model performs well in detecting hateful users but has problems in distinguishing normal users from hateful ones. This is seen in the larger number of false positives in the matrix's upper right corner, compared to false negatives in its lower left corner. 

An in-depth error analysis is conducted to further explore the types and sources of the model's misclassification. For both false positives and false negatives, we first identify quantitative differences in user and network-level features between correctly classified users and their misclassified counterparts for the whole dataset. In a second step, in a similar vein to \cite{Vidgen2020}, we use the inductive, data-derived qualitative approach of grounded theory to create a typology of misclassified users. Grounded theory is well suited for error analysis as it induces themes from the data, organises them into categories, and iterates over this process until a `saturated' set of mutually exclusive and collectively exhaustive categories is achieved \cite{Vidgen2020}. %Grounded theory helps incorporate existing theories on the challenges of accurate and fair automated hate detection, such as skewed misclassification rates for minority-group dialects, while simultaneously allowing new categories to be induced from the data itself \cite{Canonico2018}.
For this qualitative step, we select a random sample of 20 users for both false positives and false negatives, and all their tweets (up to 200) are analysed. This is done for the entire dataset, as well as the entire subset of African-American users, reflecting the focus of each research question.

%\subsubsection{Quantitative analysis of misclassified users}

\subsubsection{False positives}
Normal users who are wrongly classified as hateful (false positives) comprise the majority of errors ($\sim$80\%). Out of these, 23 ($\sim$5\%) were suspended by Twitter a few months after the dataset's creation, indicating that there may have been an annotation error where actually hateful users were labelled as normal. However, we cannot verify this.
%For model-attributable errors, comparisons of false positives with true negatives and true positives may provide greater clarity. 

Interestingly, in many regards users in the false positive category are more similar to correctly classified hateful users, and dissimilar to correctly classified normal users. This may explain why the GraphSAGE model made errors in its classifications.
Firstly, in terms of their social network: only 3\% of users in the correctly classified normal user group (true negatives) had hateful neighbours but more than 50\% of users in the false positive group did. 
The GraphSAGE models, which rely on the networks of users, may have failed to adequately distinguish between users whose immediate networks contained a high proportion of hateful users. This indicates that the model still encodes some `guilt by association' but this is likely far less than with other methods.
Secondly, the average number of words that users in the false positive category use from ADL and Hatebase's hateful lexicon is around 16\% higher than correctly classified normal users.
Lastly, users in the false positive category display more negative sentiment in their tweets than even hateful users. 
However, there are no meaningful differences observed in other network or user-level categories such as degree distributions or activity metrics. This analysis indicates that on a network and textual level, false positive users are more similar to correctly classified hateful users than to the correctly classified normal ones. 

Using a grounded-theory approach, qualitative inspection of the false positives shows they fall into four categories: annotation errors, ambiguous content, offensive (but note hateful) content, and negative sentiment against minorities. In no cases did the GraphSAGE model make a mistake which is entirely unexplainable.

\paragraph{Annotation errors} These are cases where annotators have made errors and as such are not attributable to the machine learning models. %Such errors are common in annotation processes surrounding automated hate detection, where the subjective nature of hate combined with the individual background of the annotator can often cause erroneous judgements \cite{waseem-2016-racist}. 
Five users (25\% of the false positives we reviewed) fall into this category. They include both explicit and more subtle forms of hate. For example, one user in this category clearly expressed hate against Jews with the statement: ``Because Jewish perverts like Weinstein, Woody Allen and Roman Polanski aren't unique, they're a trend". Others, however, express hate in more subtle ways, such as by using uncommon hateful slurs. For example, 2 misogynistic users often claimed that victims of rape deserved it and used terms such as `soyboy', a deeply misogynistic term used in Men's Liberation Movement groups but relatively rare in the overall online hate landscape \cite{Jones2019}. Moreover, 4 of these 5 users expressed irritation with Twitter about moderating their tweets, for example: ``Every post I make is marked sensitive. It is Twitter's soft censorship"

\paragraph{Ambiguous content} This category includes errors where it is unclear whether hate was expressed. Some users are difficult for even a human to categorise, either because further context is needed to interpret their content or because the remarks are genuinely indeterminate. Examples of such content is uses referring to a `slut culture problem', attributing something to the `black race' where context is not present on what is being attributed, or implying connections between Muslims and terrorism. Further context is often needed in such cases to establish their hatefulness. Note that some of these `ambiguous' cases could easily be considered annotation errors by some viewers (i.e., they could be genuinely hateful); however, on balance we believe there is insufficient evidence to make this stronger claim.

\paragraph{Offence vs. Hate} Around 18\% of false positive users tweeted in an offensive but not overtly hateful manner, although the line between these two is itself often hard to establish \cite{Davidson2017}. All such users had a considerable number of tweets with offensive words or phrases such as `cunts', `pussies', `whore', `fuck off', or `faggots'. Some tweeted about minority groups in an offensive, though not explicitly hateful way. For example:  ``Being ethnic does not stop you going to Sandhurst, being as thick as pig poo does. Positive discrimination is dumbing down" or  "My ex boss would not employ women or coloureds because he thought they were lazy". 

\paragraph{Negative sentiment but not hate against minorities} Almost 30\% of all false positives expressed negative sentiment but not hate against minorities, and were possibly misclassified by the model partly due to this factor. LGBT groups, immigrants, and Muslims features prominently in such cases. Examples include: ``She's illegal you dipshit. A felon. She has no rights in this country, nor do you in hers. Gtfoh" or ``Liberals: Lets use Islamic teachings to protect women. Islam: Lets make women property. Weird world we live in".

\subsubsection{False negatives}
Hateful users who are wrongly classified as normal (false negatives) comprise the minority of errors ($\sim$20\%). 
Interestingly, in contrast with the false positives, these users are not highly dissimilar to users from their true class (here, hateful). That said, they exhibit some notable similarities with users in their wrongly predicted class (i.e., normal), which may partially explain why the model misclassified them.
A far lower proportion of false negatives had normal neighbours compared with an average hateful user: on average, around 80\% of correctly classified hateful users had a normal neighbour, compared to only 51\% of false positive users. %Both groups also had 100\% of their members having at least one hateful neighbour. 

Even on a textual level, the average number of hateful words used by the hateful users (true positives) and the false negatives is nearly equal, and 37.5\% higher than the average for normal users. Yet in other areas such as sentiment and subjectivity, false negatives are systematically dissimilar to hateful users. False negatives show, on average, 25\% greater positive sentiment in their tweets versus hateful users.

%\subsubsection{Qualitative analysis of misclassified users' tweets using grounded theory}
%\paragraph{False negatives}

Qualitative analysis of false negative users reveals three major categories: annotation errors, hate against uncommon targets, and non-explicit hate.

\paragraph{Annotation errors} A large portion of false negatives (40\%) were due to annotation errors.
Firstly, hateful labels were often wrongly assigned to users who were, in fact, opposed to hateful groups and used strongly-worded language to defend minority groups. For example, a user who was against the genocide of Rohingya Muslims in Myanmar and expressed vigorous opposition to its regime was annotated as hateful.
Secondly, a quarter of these wrongly annotated users used either obscene or offensive language, but in self-referential and non-hateful ways. 
Lastly, a third of these annotation errors were for African-American users. They showed frequent use of words such as `nigga', `bitch', or  `ass' but always in non-hateful ways and likely in community contexts. Such errors reflect previous findings that even annotators focus strongly on words only, often ignoring the context of language use and thus mislabelling users as hateful \cite{Mozafari2019}. Note that no such errors were identified with African American users for the false positives.

\paragraph{Hate against uncommon targets} Of the remaining 60\% of false negatives that can be directly attributed to the model, a quarter contain users who express hate against uncommon targets. Such users were indeed hateful, but directed their hate against Native-Americans, old-aged people, or Christians. These categories of victims are uncommon on social media: research by \citeA{Mondal2017} on categorising hate speech on Twitter by its targets find that hate directed against religion or ethnicity make up less than 2\% of all hate speech. Hate against old people does not feature at all in their data. Given the scarcity of such examples, ML models may not adequately learn to identify such forms of niche hate.

\paragraph{Implicit hate} 
A third of the model-attributable false negatives were for implicit hate: hateful users who did not explicitly use hateful words or phrases in their texts but relied on contextual references. For example, two users praised ISIS's actions and someone who ``beat up the black guy on the bus in that video" respectively. Another favourably referenced what Hitler would do in a particular situation to express racist sentiment. This category also featured users who use subtle but still hateful words when diminishing women, such as `broad'. Such cases are generally more difficult for annotators to label.

\subsubsection{Errors among African-American users}
For African-American users the GraphSAGE model only produced false negatives, with no false positives.
50\% of these were annotation errors where words, phrases and other linguistic features common to AAVE were likely mistaken as signals for the hateful nature of a user, a source of annotation error common in many other popular hate detection datasets \cite[see][]{Dixon2018, Sap2019}.
Users with tweets such as ``That's flint nigga not Detroit. Dumbass" or ``God damn your bitch ass backup account do numbers like this?" were mislabelled as hateful.
From an analysis of all their available tweets, however, it is clear that they use offensive or otherwise hateful terms such as `nigga' or `bitch' in community contexts only, without any overt intention of harming a potential target.
Out of the other half, which comprised errors produced by the model, one user had been suspended from Twitter. They expressed anti-LGBT hate, albeit more subtle varieties. The others either had content which could be characterised as offensive, for example ``I like nasty bitches", or expressed hate implicitly.

\section{Discussions}\label{sec:discussion}
% and Limitations

\subsection{Towards more accurate classification}

\subsubsection{The promise of networks}
Our results show that the incorporation of dynamically learnt network representations along with user-level and textual information through geometric deep learning produces more accurate classification of hateful users. Across all metrics, GraphSAGE performs the best. Manual engineering of such network information still helps boost accuracy over techniques not including such features but to a lesser extent.

From a social science perspective, these results help support existing theories about homophily among hateful users by demonstrating how network-level insights can be leveraged successfully to boost classification accuracy.
However, the most accurate results are only achieved when users' network-related information is dynamically learnt; the Recall for the ANN incorporating static network features is actually the lowest among all models.
Social psychology theories of group dynamics imply that normal users should be less connected to hateful ones: clustered into communities, hateful users benefit from the `justification forces' of each other's approval while seeking to minimise `suppression forces' provoked by disapproval of their views or defence of their targets \cite{Day2018}.
There was only weak evidence for this from the retweet network, with normal users still having a considerable proportion of hateful neighbours. Even so, models including network information have the highest Precision scores, indicating that they are best able to distinguish between the two classes of users by using their networks. Leveraging such features more efficiently and effectively could demonstrably improve flagging of hateful users in the real world.

\subsubsection{Limits on accurate hateful user classification}
Despite the advances that Geometric deep learning offer, Precision remains fairly low overall, with the best classifier only having a score of 46.1. While this may be because of the nature of the retweet network, where normal users do not show particularly strong assortative mixing, it may also arise due to training data scarcity, a topic discussed in further detail below. Furthermore, as the error analysis demonstrates, annotation errors comprised a substantial proportion of both false negatives and false positives and could also contribute to this, artificially depressing the model's performance. Any such analysis would have to also account for the underlying tensions faced by human annotators in labelling such content. As \citeA{Vidgen2020} write, for ``complex and often-ambiguous content even well trained annotators can make decisions which are inappropriate". Disagreement metrics between annotators, such as Kappa scores, are essential for understanding whether the model's errors are related to annotator uncertainty (i.e., the `ground-truth' label is contested in the first place). Unfortunately, no such metrics were collected by the datasets' creators, preventing such analysis in this case.

The model performs poorly when users' content is ambiguous or where hate is not explicitly expressed. It is  unable to adequately distinguish between offence and hate, possibly because it focuses on hateful words without fully understanding the nuanced context in which they are used. This is partly because of the nature of the data itself. Tweets are short-form content, often implicitly reference other entities, and can be ambiguous, even when considered by humans \cite{MacAvaney}. A way to ameliorate this problem may be through the consideration of heterogeneous graphs, where users and their tweets are considered as heterogeneous but connected nodes, and different types of edges are placed between users and between their tweets if a tweet references another one (for example, through a retweet or reply). Indeed early experiments by Facebook AI in using heterogeneous graphs combined with graph convolutional networks for hateful \textit{content} detection have shown promising results \cite{Mishra2019-gcn}. Extensions of this could fruitfully be made to hateful \textit{user} detection. Such efforts may help models better resolve the inherent ambiguity in short-form content by considering the wider conversational context of each tweet.

Another way to address this problem is through using a better representation of each users' content. We used GloVe embeddings (provided by the original authors) but contextual word embeddings, such as BERT, have been shown to achieve better performance in classifying hateful content \cite{Mozafari2019}. Unfortunately, it was not possible to generate such embeddings for the data at hand as the content of several users was missing in the dataset. Thus, while geometric deep learning may help resolve certain classification challenges by incorporating network information, its impact is still limited overall. Further research is needed into ambiguous content to fully understand the limits of incorporating social network information.

\subsection{Geometric deep learning: A fairer approach?}

Alongside boosting accuracy, incorporating learnt network representations of users into the classification task also reduces longstanding biases against African-Americans in automated hate detection. This boosts fairness as far as `predictive equality' is concerned, where non-discrimination towards the minority group is required only for a particular outcome of interest, which in this context is being classified as hateful.
Without the inclusion of network information, the ANN classifier produced disparate impact on African-American users, who had an FPR almost 6 percentage points higher than users from other demographic groups. The inclusion of network information within the ANN itself helped somewhat to equalise the FPRs between the minority and majority groups.
When such network representations are learnt then the FPR declines further for the African-American group, dropping to zero for GraphSAGE with mean aggregation, which is also the most accurate classifier overall. Thus, it appears that the selection of the `predictive equality' fairness criteria, which unlike other criteria such as `demographic parity' does not place strict limits on classification accuracy, has the desired effect: the fairest classifier is also the most accurate one.
%Moreover, the `fairness-accuracy trade-off' may also be mitigated by the choice of model itself, with a more context-aware model being better able to accurately classify previously disadvantaged groups.

These results, however, should be interpreted with some caution. Firstly, the subset of African-American users identified is small, and testing on a larger group is required to fully establish the fairer nature of geometric deep learning classifiers. Secondly, the annotation errors observed in the error analysis complicate what conclusions can be drawn from the results. Half of the African-American users labelled as hateful are, in fact, normal. 
%It may also be possible that some African-American users who were labelled as normal were, in fact, hateful. A new annotation process, where users are relabelled in a less biased manner, may help further determine the validity of the model's results. 

Additionally, geometric deep learning methods decrease the FPR among African-American users to a rate that is \textit{lower} than that of the non-African-American group. From a purely technical perspective, where `predictive equality' is concerned, the FPR between the two groups should be close to, if not exactly, equal. The large gap in the FPR rates, despite addressing historical bias against a minority group, thus poses a problem from this viewpoint. From a political and moral philosophy perspective, judgements on fairness are often driven by broader normative principles \cite{Leben2020}. For example, ``there may be a greater social cost associated with keeping peaceful black prisoners in jail than peaceful white prisoners, since this perpetuates a historical cycle of deprivation in the black community" \cite{Leben2020}. A very similar case arises in the context of the current research. As described above, there has been significant past bias in hate detection algorithms against African-Americans, whose speech has been persistently suppressed. As such, a normative decision is made to assign greater weight to the FPR for this group than for the non-African American group. This is arguably more reflective of societal ideals of fairness.
%Such normative considerations, once acknowledged and adjusted for, render the outcomes of geometric deep learning techniques `fair', despite their different FPR between African-American and non-African-American groups. 

\subsection{Data scarcity as a limiting factor}

The performance of supervised machine learning techniques, as used here, is highly dependent on the number of labelled samples from which the models can learn. In this regard, both the limited size of the annotated dataset and the  extremely low number of hateful users among them is problematic. Hateful users, numbering at 544 only, make up less than 10\% of the already small dataset of 4,972 annotated users. For comparison, hate detection research focused on content classification often use datasets with more than 20,000 labels \cite[e.g.,][]{Vidgen2020, Davidson2017}. Thus, while even the best performing model used has relatively low precision and F1-score, it may be that adding more annotated data for training may improve the results. Indeed, when increasing the training dataset ratio to 90\% from the 80\% used in 5-fold cross validation, the performance of GraphSAGE with mean aggregation improves across all metrics.

%Data scarcity also presents a problem where the evaluation of fairness is concerned. The reliability of the demographic detection technique is discussed in further detail below. However, if its results are taken at face value, then the number of African-American users in the dataset is very small. With only 136 such users, an evaluation of model performance on this subset may not be indicative of its overall fairness. The model's results may be too specific to this tiny subset to generalise to African-American users as an entire group. 
% Moreover, the size of the dataset also implies that very small perturbations in the model's performance could have a major impact on the FPR and therefore the evaluation of model fairness. 
Thus, while promising, further data is needed to make more robust conclusions about the extent to which dynamically learnt network information through geometric learning can increase the accuracy and fairness of automated hateful user detection.

\subsection{The reliability of demographic inference}
Demographic attribute inference from individuals' online digital traces is difficult, especially where race is concerned \cite{Hinds2018}. This hurdle is also faced in this research with respect to the reliability of the methods used to identify identify users as African-American.  The LDA topic model developed by \cite{Blodgett2016} and used here was tested further on separate datasets; it achieved strong results compared to other methods in accurately identifying African-American users \cite{Preot2018}. Even so, it achieved an overall AUC of 0.729, meaning that errors still persist in the model's predictions \cite{Preot2018}. While such errors may have been mitigated to an extent by the second step of manually verifying each identified user, even such a two-step approach can cause significant issues. 
Firstly, the LDA model's predicted were exclusively driven by the textual information of each user. Yet, \citeA{Preot2018} and \citeA{Wang2019} have demonstrated that detection of a user's demographic traits can be significantly boosted if their profile information, such as profile image, is used in addition to textual content. 
Secondly, even when a second step of manual verification was conducted to mitigate against the model's performance concerns, only the text of most users could be examined as the original dataset of tweets was incomplete and some accounts were no longer active. Thus, it is likely that the lead author's own judgement may have led to a wrong race being assigned to a particular user, as textual information can be interpreted in myriad ways, besides being inadequately informative by itself.
Thirdly, the additional verification could only be performed on users detected by the model at a first step as African-American. While this may help reduce the possibility of false positives, no error mitigation was done where false negatives---that is, users who are African-American but not identified by the model as such---are concerned.
Lastly, as \citeA{Blodgett2016} themselves acknowledge, individuals may deploy different dialects when communicating online. Deciphering demographics by using dialect as a proxy is therefore itself likely to lead to errors, despite fairly high accuracy when tested. 
%In this regard, the application of the LDA model to only up to 200 tweets per user over a limited time period is problematic: it may lead the model to identify the only the primary dialect of a user in a specific time window, and wrongly using this as a proxy for their ethnicity.
A more systemic manual annotation of users in this dataset based on their race would thus be beneficial in helping draw stronger conclusions about the ML models' outcomes.

\section{Conclusion}\label{sec:conclusion}

We have demonstrated the promise of network-level information, learnt through geometric deep learning, for achieving fairer and more accurate automated hate detection. We have also outlined the need for greater focus on hateful user detection (rather than hateful content detection) and contributed to this nascent area of research. At the same time, we have highlighted the potential for bias in this task and its negative implications on falsely classified users. It is one of the first works (to our knowledge) to both consider the problem of fairness in hateful user detection and to develop a model that minimises it.

To complete this work we have leveraged both social science theories and advances in ML research. The task of automated hate detection is, despite the widespread use of ML models, an inherently social one. Combining social science insights on homophily between hateful users with geometric learning techniques, this paper has enabled creation of a more effective classifier. It has evaluated this classifier, moreover, by choosing an appropriate notion of fairness that is guided by broader concerns in society about historic algorithmic bias against African-Americans (predictive equality). We have also argued that this notion of fairness may itself face problems in cases where the FPR can be minimized for historically discriminated-against groups more than other groups; this may not technically optimize the FPR but accords better with social ideals of fairness. 

In sum, this paper has taken a step in combining social science insights with ML advances to produce more context-aware, accurate and fairer automated hate detection systems, designed for users rather than content. With this contribution, it hopes to inspire future work in this nascent but important area of research.

\bibliographystyle{ACM-Reference-Format}
\bibliography{references}

\end{document}